\documentclass[aip,jcp,amsmath,reprint]{revtex4-1}
\usepackage{graphicx}

\begin{document}

\title{Comment on: ``Static correlations functions and domain walls in
  glass-forming liquids: The case of a sandwich
  geometry'' [J. Chem. Phys. 138, 12A509 (2013)]}

\author{Vincent Krakoviack}
\affiliation{Laboratoire de chimie, {\'E}cole normale sup{\'e}rieure
  de Lyon, 46 all{\'e}e d'Italie, 69364 Lyon cedex 07, France}

\date{\today}

\begin{abstract}
  In this Comment, we argue that the behavior of the overlap functions
  reported in the commented paper can be fully understood in terms of
  the physics of simple liquids in contact with disordered substrates,
  without appealing to any particular glassy phenomenology. This
  suggestion is further supported by an analytic study of the
  one-dimensional Ising model provided as Supplementary Material.
\end{abstract}

\maketitle

In a recent paper,\cite{GraTroCavGriVer13JCP} Gradenigo et al.\@ have
reported on a computer simulation study of a glass-forming liquid
constrained by amorphous boundary conditions representative of its
equilibrium bulk configurations. This setup has been put forward a few
years ago as a possible tool to probe the existence of nontrivial
static correlations in bulk glassy systems, through the investigation
of point-to-set correlation functions such as configurational
overlaps.\cite{BouBir04JCP,MonSem06JSP_2} Accordingly, quantities of
this type are reported in Ref.~\onlinecite{GraTroCavGriVer13JCP} for
two related geometries, a semi-infinite fluid in contact with a single
wall (wall geometry) and a fluid slab sandwiched between two parallel
walls (slit geometry). These data are then analyzed in terms of the
interplay between the boundary conditions and the complex
coarse-grained free-energy landscape postulated by the random
first-order transition (RFOT) theory for bulk glassy
liquids.\cite{BouBir04JCP,CamBirTarTar11PRL,BirCam14Arxiv}

However, it was recently observed\cite{Kra14JCP} that, outside the
realm of glassy physics, constrained systems such as those of
Ref.~\onlinecite{GraTroCavGriVer13JCP} are just special instances of a
generic model for fluids in contact with random substrates previously
studied with standard tools of the theory of simple
liquids.\cite{DonKieRos94PRE} In this framework, it is customary to
quantify the direct influence of the quenched-disordered solid
boundary on the microscopic fluid configurations via the so-called
blocking or disconnected two-point density correlation function, i.e.,
the covariance of the random density profile established in the
presence of the amorphous surface. This correlation function can be
straightforwardly turned into configurational overlaps that are
analogues of those measured in Ref.~\onlinecite{GraTroCavGriVer13JCP}
and not necessarily bound to be featureless objects, even in the
absence of specific glassy features.\cite{ChaChaTar13JCP,Kra14JCP}

From this observation, the question naturally arises, whether one
really needs to appeal to any particular glassy phenomenology to
interpret the behavior of the overlap functions reported in
Ref.~\onlinecite{GraTroCavGriVer13JCP}. This is the point addressed in
this Comment, through a direct comparison between the wall and slit
geometries. For reference, it should be recalled that, in a
RFOT-inspired analysis, the two geometries are expected to be ruled by
different physics, resulting in well distinct characteristic
lengthscales,\cite{CamBirTarTar11PRL,BirCam14Arxiv} and this is how
the data are described in Ref.~\onlinecite{GraTroCavGriVer13JCP}.

The present discussion is guided by an asymptotic result derived in
Ref.~\onlinecite{Kra14JCP} for the disconnected total correlation
function of a nonglassy liquid, $h_\text{dis}(\mathbf{x},\mathbf{y})$,
when at least one of the points $\mathbf{x}$ or $\mathbf{y}$ is far
enough from any amorphous boundary. Indeed, one then gets
\begin{equation}\label{eq:asymhet}
  h_\text{dis}(\mathbf{x},\mathbf{y}) \simeq \rho \int d\mathbf{u} \,
  h(|\mathbf{x}-\mathbf{u}|) \chi(\mathbf{u})
  h(|\mathbf{u}-\mathbf{y}|),
\end{equation}
with $\rho$ the number density of the fluid, $h(r)$ its total
correlation function in the bulk, and $\chi(\mathbf{r})$ the indicator
function of the domain from which the fluid particles are excluded by
the amorphous boundaries.

The linearity of this equation with respect to $\chi(\mathbf{r})$
suggests to investigate the range of validity of a simple
superposition approximation, in which, for the geometries considered
here, the effect on the fluid of the two walls in the slit geometry
would merely be the sum of the effects of the two walls taken
individually. Note that such a linear regime can be expected on
general grounds and that Eq.~\eqref{eq:asymhet}, which represents the
leading asymptotic contribution to it, only plays the role of a formal
proof of existence. In terms of the configurational overlaps reported
in Ref.~\onlinecite{GraTroCavGriVer13JCP}, such a superposition
approximation leads to the compact relation
\begin{equation}\label{eq:overlaps}
  q_c(d)-q_0 \simeq 2[q(d)-q_0],
\end{equation}
with $q_c(d)$ the overlap at the central plane of a slit of width
$2d$, $q(d)$ the overlap at a distance $d$ from a single wall, and
$q_0$ the trivial ideal-gas contribution to these functions.

\begin{figure}[t!]
\includegraphics[scale=0.92]{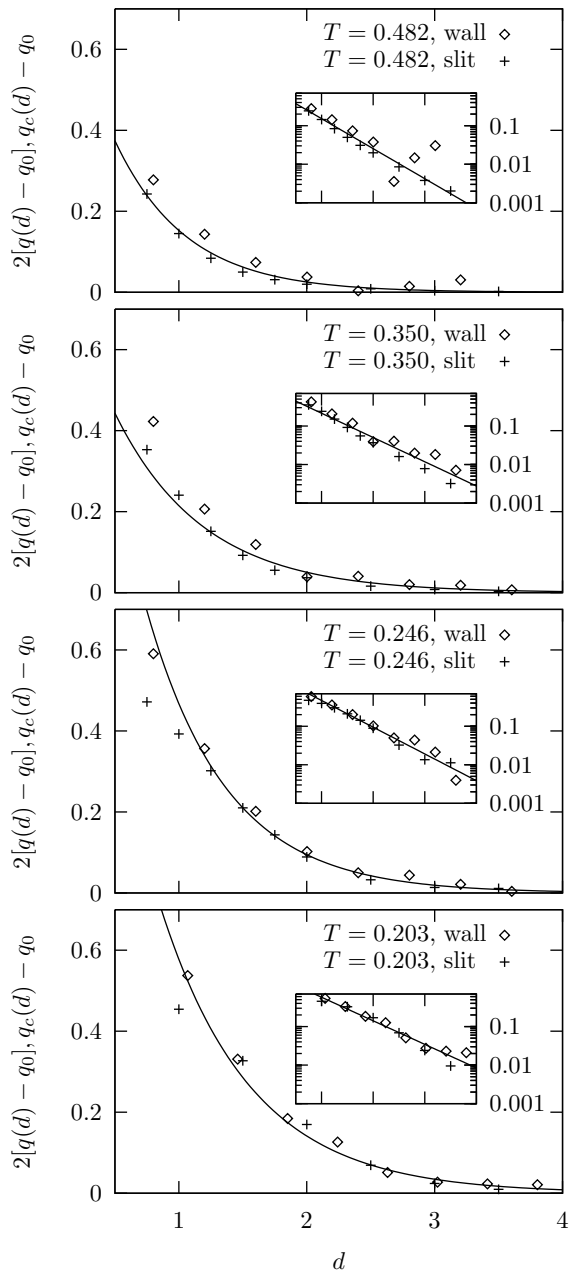}
\caption{\label{figure} Comparison at four temperatures of the excess
  overlap $q_c(d)-q_0$ at the central plane of a slit of width $2d$
  with twice the excess overlap $q(d)-q_0$ at a distance $d$ from a
  single wall. The continuous lines are joint exponential fits for
  $d \geq 1$. (Insets) Same data in semi-log plots.}
\end{figure}

Equation \eqref{eq:overlaps} is tested at four temperatures in
Fig.~\ref{figure}, where the data of
Ref.~\onlinecite{GraTroCavGriVer13JCP} are plotted accordingly. It
appears quite reasonable at all temperatures for $d \gtrsim 1$ and, in
this domain, $2[q(d)-q_0]$ and $q_c(d)-q_0$ can be both described by
the same exponential decay. This suggests that, in this regime in both
geometries, the behavior of the system is ruled by rather simple
amorphous-boundary effects.

Gradenigo et al.\@ have shown that $2[q(d)-q_0]$ keeps an exponential
behavior down to $d=0$ at all studied temperatures, while $q_c(d)-q_0$
remains exponential down to $d=0.75$ (the smallest value considered
for the slit geometry) at the two highest temperatures and tends to
flatten for $d \lesssim 1$ at the two lowest. Such a bending is a
foreseeable consequence of the presence of two nearby facing
boundaries, i.e., of confinement: For narrow slits, the combined
action of the two amorphous walls is indeed expected to restrain the
decay of the correlations of the disorder-induced fluid density
profile when $d$ increases more strongly than a mere linear
superposition of independent boundary effects. In fact, such a
leveling-off disrupting at short distances a medium-to-long-range
exponential decay occurs in models as simple as the one-dimensional
Ising model, as shown in the Supplementary Material. \footnote{See
  supplementary material at [URL will be inserted by AIP] for an
  analytic study of the one-dimensional Ising model.} If no bending is
seen at the highest temperatures, this might only mean that the
breadth of the raw confinement effect, expected to decrease as
temperature increases, is too small at these temperatures for it to
appear in the probed slit-width window.

Therefore, by looking at the data of
Ref.~\onlinecite{GraTroCavGriVer13JCP} from the angle of the physics
of simple liquids in contact with disordered substrates, i.e., of raw
boundary and confinement effects, \cite{DonKieRos94PRE,Kra14JCP} we do
not see any compelling evidence that any specific glassy phenomenology
has to be appealed to in order to give an account of the observed
behaviors. In particular, in the studied temperature range, there is
no obvious sign of distinct and complex physics in the wall and slit
geometries, both described by the same simple exponential decay law
starting at quite small distances already, at variance with
expectations from the RFOT scenario, for
instance. \cite{GraTroCavGriVer13JCP,CamBirTarTar11PRL,BirCam14Arxiv}

The contrasting conclusions of Ref.~\onlinecite{GraTroCavGriVer13JCP}
and of this Comment clearly point towards difficulties in the
interpretation of measured point-to-set correlation functions. They
actually are complex objects, possibly blending ingredients from the
physics of normal and glassy liquids that are not easily sorted
out. \cite{Kra14JCP} In fact, these difficulties were already
acknowledged in Ref.~\onlinecite{GraTroCavGriVer13JCP}, where warnings
were raised, based on the experience with an alternative setup, the
so-called random pinning geometry.\cite{SzaFle13EPL,ChaChaTar13JCP}
However, they could not be made concrete, due to the lack of simple
results such as Eq.~\eqref{eq:asymhet} that appeared more recently.

Finally, it should be mentioned that the predictions of the RFOT
theory for the point-to-set correlations are one aspect of an
elaborate scenario, also involving features such as a bimodal
distribution of overlaps recently observed in computer simulations of
spherical cavities. \cite{BerChaYai16JCP} It remains a challenge for
the future to find out whether the simple picture put forward in this
Comment could also account for these additional aspects.

\bigskip

The authors of Ref.~\onlinecite{GraTroCavGriVer13JCP} are warmly
thanked for sharing their data with us and thus making the present
analysis possible.


%

\clearpage

\noindent\textbf{\sffamily \large Supplementary Material}
\medskip
\setcounter{equation}{0}
\setcounter{figure}{0}

In this Supplementary Material, we provide and discuss the analytic
results for the spin overlap functions of the zero-field
one-dimensional Ising model in the ``wall'' and ``slit''
geometries. The calculation is a simple exercise in the application of
the standard transfer matrix method, whose main elements of solution
can be found in the work of Grinstein and Mukamel on a special
instance of the random-field Ising model.\cite{GriMuk83PRB}

\paragraph*{Preliminary.}
We consider the standard Ising Hamiltonian for $N+1$ spins,
\begin{equation}
  \beta H[\mathbf{S}] = - \beta J \sum_{i=0}^{N-1} S_i S_{i+1},
\end{equation}
with $S_i=\pm 1$, $i=0,1,\ldots,N$. $J>0$ is the exchange constant and
$\beta$ the inverse temperature. For latter use, we define
$\tau=\tanh(\beta J)$.

Assuming that the spins $S_0$ and $S_N$ are frozen, one gets the
conditional thermal averages,\cite{GriMuk83PRB}
\begin{gather}
  \langle S_k \rangle^{S_0 S_N}_N = S_0 \frac{\tau^k + S_0 S_N
    \tau^{N-k}}{1+S_0 S_N \tau^N}, \\
  \langle S_k S_l \rangle^{S_0 S_N}_N = \frac{\tau^{l-k} + S_0 S_N
    \tau^{N-l+k}}{1+S_0 S_N\tau^N}, \ k\le l.
\end{gather}

\paragraph*{Bulk behavior.}
Taking $N\to\infty$ and $0\ll k\le l\ll N$, one gets for the spin
correlation function,
\begin{equation}
  \langle S_k S_l \rangle_\infty = \tau^{l-k} = e^{(l-k)\ln\tau}.
\end{equation}
The $S_0$ and $S_N$ dependence vanishes as it should, and the bulk
correlation length follows as $\xi_\text{b}=-1/\ln\tau$.  

From this correlation function, one can straightforwardly derive the
probabilities that two distant spins are parallel or anti-parallel in
the bulk,
\begin{equation}
  P_{S_k=S_l} = \frac{1+\tau^{l-k}}{2},\ P_{S_k=-S_l} =
  \frac{1-\tau^{l-k}}{2},\ k\le l. 
\end{equation}

\paragraph*{``Wall'' geometry.}
Taking $N\to\infty$ and $k=L \ll N$, the average magnetization for
fixed $S_0$ reads
\begin{equation}
  \langle S_L \rangle^{S_0}_\infty = S_0 \tau^L,
\end{equation}
independently of $S_N$.  

Squaring this result and performing a realization average over the
value of $S_0$ (this is needed in principle, but immaterial for this
specific calculation), one obtains the spin overlap at a distance $L$
from the ``wall'', i.e., spin $0$,
\begin{equation}
  Q(L) = \overline{[\langle S_L \rangle^{S_0}_\infty]^2} = \tau^{2L} =
  e^{-2L/\xi_\text{b}},
\end{equation}
where $\overline{\cdots}$ denotes the disorder average.

\paragraph*{``Slit'' geometry.} 
Taking $N=2L$ and $k=L$, the average magnetization for fixed
$S_0$ and $S_{2L}$ reads
\begin{equation}
  \langle S_L \rangle^{S_0 S_{2L}}_{2L} = S_0 \frac{(1 + S_0 S_{2L})
    \tau^L}{1+S_0 S_{2L} \tau^{2L}}.
\end{equation}
Squaring this result and performing a realization average over the
values of $S_0$ and $S_{2L}$ with the probabilities $P_{S_0=S_{2L}}$
and $P_{S_0=-S_{2L}}$ (this is the condition to have frozen boundaries
representative of the equilibrium bulk configurations), one obtains
the spin overlap at the center of a ``slit'' of width $2L-1$ delimited
by spins $0$ and $2L$,
\begin{equation}
  Q_c(L) = \overline{[\langle S_L \rangle^{S_0 S_{2L}}_{2L}]^2} =
  \frac{2 \tau^{{2L}}}{1+\tau^{2L}} = \frac{2
    e^{-2L/\xi_\text{b}}}{1+e^{-2L/\xi_\text{b}}}.
\end{equation}
 
\paragraph*{Discussion.}  %

The one-dimensional Ising model is a minimalist version of the problem
at hand, a caricature, actually. It clearly lacks ingredients that can
be expected to play an important role in the fluid-based systems, such
as the roughness of the boundaries, which is lost because of both its
one-dimensional character and lattice-based structure. An
oversimplified form of disorder results, which is of an effectively
binary nature and fully encoded in the variable $S_0 S_{N}=\pm
1$. Remarkable properties are observed, such as the perfect scaling
behavior with respect to the bulk correlation length $\xi_\text{b}$,
that cannot necessarily be expected from fluid systems at the particle
scale. Yet, the model provides an interesting opportunity to
demonstrate the effects of the wall and slit constraints on an
otherwise almost featureless system.

\begin{figure*}
\hfill \includegraphics[scale=0.8]{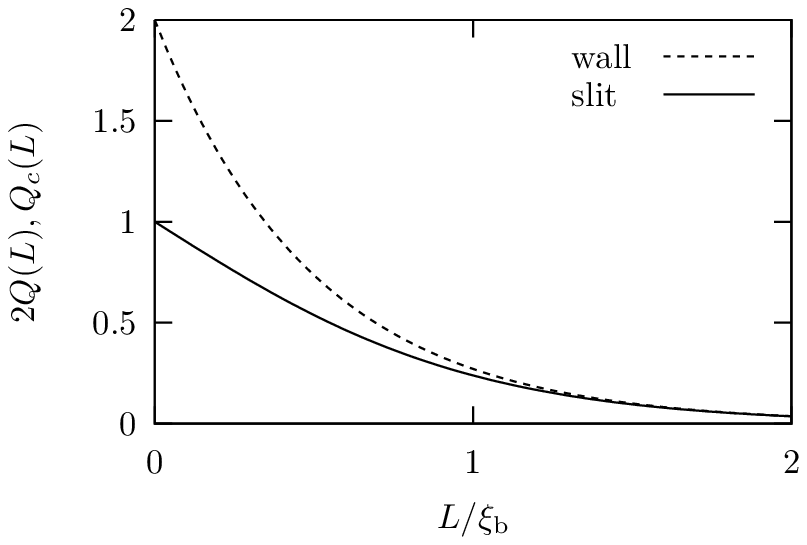} \hfill
\includegraphics[scale=0.8]{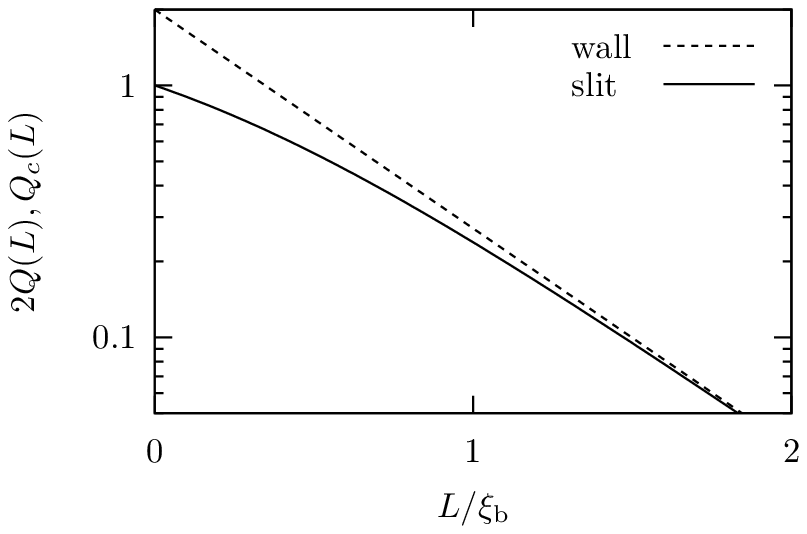} \hfill\null
\caption{\label{suppfig1} Linear and semi-log plots of the spin
  overlaps of the one-dimensional Ising model in the ``wall'' and
  ``slit'' geometries. The distances are scaled by the bulk
  correlation length $\xi_\text{b}$.}
\bigskip\bigskip\medskip

\hfill \includegraphics[scale=0.8]{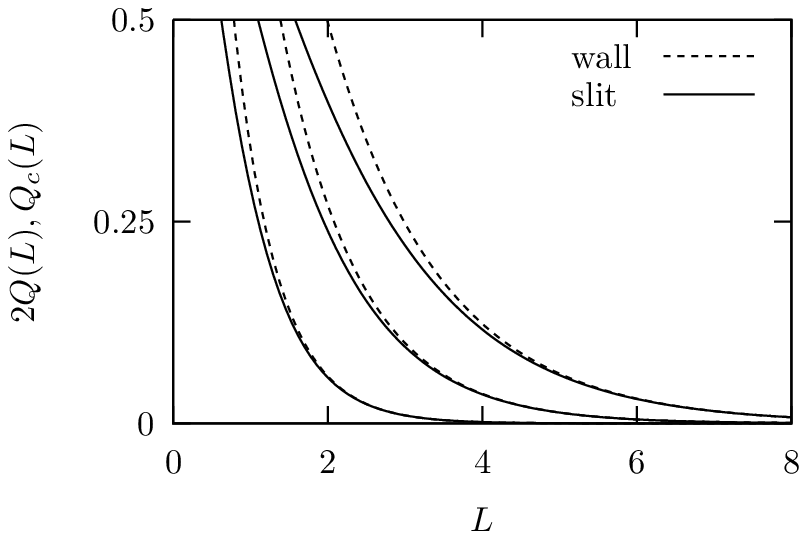} \hfill
\includegraphics[scale=0.8]{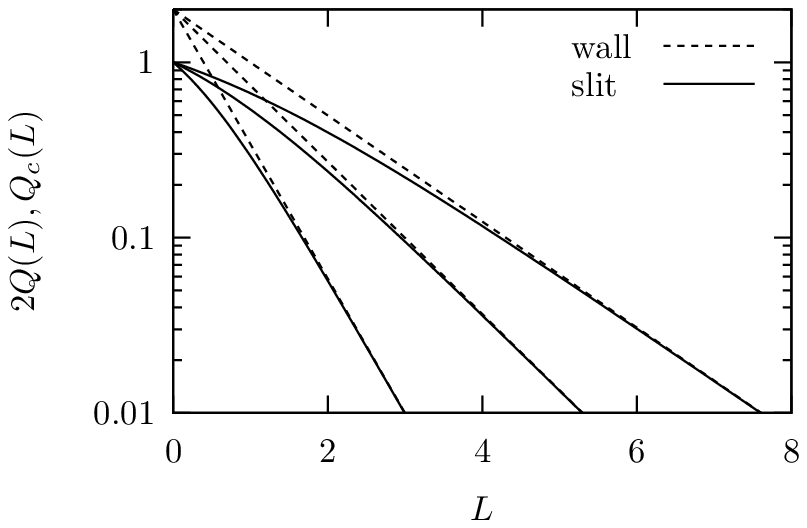} \hfill\null
\caption{\label{suppfig3} Linear and semi-log plots of the spin
  overlaps of the one-dimensional Ising model in the ``wall'' and
  ``slit'' geometries at different temperatures. From left to right,
  $T=\beta^{-1}=1.6 T_0$, $T_0$, $0.8 T_0$, where $T_0\simeq 1.42 J$
  is such that the bulk correlation length $\xi_\text{b}(T_0)=2$.}
\bigskip\bigskip\medskip

\hfill \includegraphics[scale=0.8]{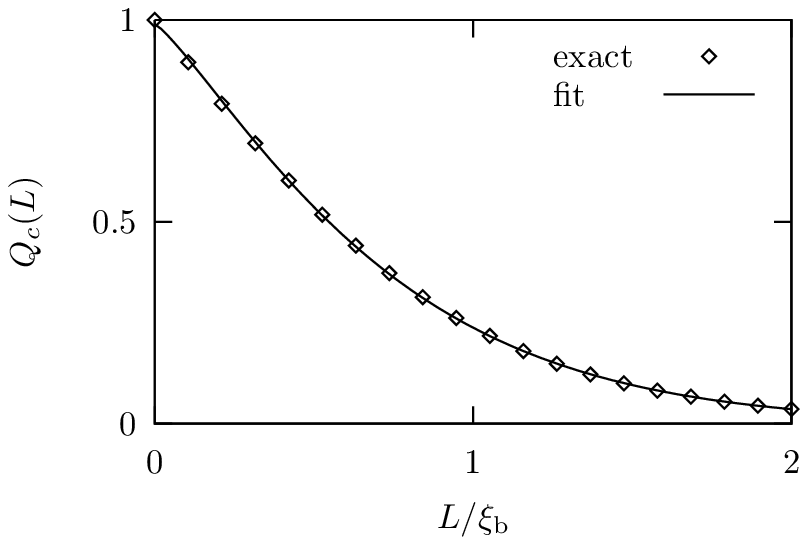} \hfill
\includegraphics[scale=0.8]{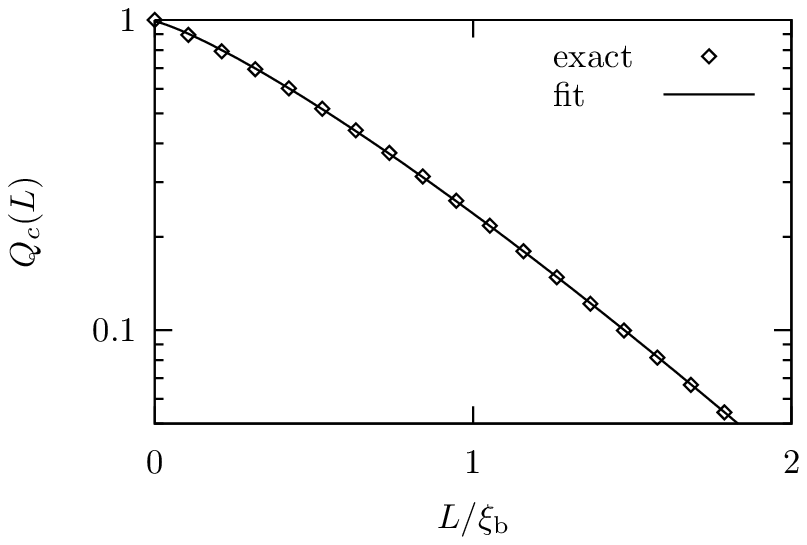} \hfill\null
\caption{\label{suppfig2} Linear and semi-log plots of the discretely
  sampled spin overlap of the one-dimensional Ising model in the
  ``slit'' geometry and its best fit by a compressed exponential law
  on the domain $0\leq L/\xi_\text{b} \leq 2$. The distances are
  scaled by the bulk correlation length $\xi_\text{b}$.}
\end{figure*}

The spin overlaps $Q(L)$ and $Q_c(L)$ are plotted as functions of the
scaled distance $L/\xi_\text{b}$ in Fig.~\ref{suppfig1}. They clearly
display behaviors in line with the qualitative analysis posited in the
Comment for the overlap functions of glass-forming liquids. Indeed,
for $L/\xi_\text{b}$ large enough, the asymptotic relations
\begin{equation}
  Q_c(L) \simeq 2 Q(L) = 2 e^{-2L/\xi_\text{b}} 
\end{equation}
hold. Moreover, $Q(L)$ is found to be strictly exponential all the way
down to $L/\xi_\text{b}=0$, while $Q_c(L)$ levels off at small
$L/\xi_\text{b}$ as a consequence of the confinement effect in narrow
slits.

It is also interesting to consider the explicit temperature evolution
of $Q_c(L)$, whose nonexponential domain has to broaden in absolute
units of length as the temperature is lowered, because of the
associated growth of $\xi_\text{b}$. It is reported in
Fig.~\ref{suppfig3}, where one can see $Q_c(L)$ crossing over from an
exponential to a nonexponential shape (in the window $L\ge 2$, for
instance) as the temperature decreases. Note that the parameters of
the figure have been chosen to be in rough agreement with those
describing the situation in the glass-forming systems: There is a
factor of two between the largest and the smallest temperatures, and
the intermediate one corresponds to a microscopic value of the
characteristic decay length of $Q(L)$, $\xi_\text{b}/2 = 1$ lattice
spacing.

Taking advantage of the exact result for $Q_c(L)$, the numerical data
analysis developed in Ref.~\onlinecite{GraTroCavGriVer13JCP} can be
repeated, by fitting a coarsely sampled $Q_c(L)$ in its nonexponential
regime to a compressed exponential law
\begin{equation}
Q_c(L) \simeq A \exp[-(L/\xi)^\zeta].
\end{equation}
As seen in Fig.~\ref{suppfig2}, an excellent fit is obtained with a
moderate anomalous exponent $\zeta\simeq 1.22$ and an effective decay
length $\xi \simeq 0.75 \xi_\text{b}$, larger than the actual one
which is the same as in the wall geometry and equal to
$\xi_\text{b}/2$.

Interestingly, in the study of glass-forming liquids in the slit
geometry, a crossover from a high-temperature exponential to a
low-temperature nonexponential behavior, associated with
numerically-determined larger correlation lengths than in the wall
geometry, is similarly observed, but with larger and
temperature-dependent anomalous exponents and lengthscale ratios.  In
this context, it is interpreted as a signature of glassiness. This
obviously does not apply to the present model, in which the somewhat
weaker quantitative effects might be the mere consequences of its
rather impoverished physics.

\end{document}